\documentclass[conference]{IEEEtran}
\usepackage[letterpaper, left=0.63in, right=0.63in, bottom=1.01in, top=0.75in]{geometry}
\IEEEoverridecommandlockouts
\usepackage[binary-units]{siunitx}
\usepackage[caption=false,font=footnotesize]{subfig}
\usepackage{tabularx}
\usepackage{multirow}
\usepackage{color, colortbl}
\usepackage[binary-units]{siunitx}
\DeclareSIUnit{\bps}{bps}
\usepackage{amsmath,amssymb,amsfonts}
\usepackage{cite}
\usepackage{graphicx}
\begin{document}
\title{User Scheduling and Passive Beamforming\\ for FDMA/OFDMA in Intelligent Reflection Surface}

\author{\IEEEauthorblockN{Wei Jiang\IEEEauthorrefmark{1}\IEEEauthorrefmark{2} and Hans D. Schotten\IEEEauthorrefmark{2}\IEEEauthorrefmark{1}}
\IEEEauthorblockA{\IEEEauthorrefmark{1}Intelligent Networking Research Group, German Research Center for Artificial Intelligence (DFKI)\\
  }
\IEEEauthorblockA{\IEEEauthorrefmark{2}Department of Electrical and Computer Engineering, Technische Universit\"at (TU) Kaiserslautern\\
 }
}


%


\maketitle

\begin{abstract}
Most prior works on intelligent reflecting surface (IRS) merely consider point-to-point communications, including a single user, for ease of analysis. Nevertheless, a practical wireless system needs to accommodate multiple users simultaneously. Due to the lack of frequency-selective reflection, namely the set of phase shifts cannot be different across frequency subchannels, the integration of IRS imposes a fundamental challenge to frequency-multiplexing approaches such as frequency-division multiple access (FDMA) and the widely adopted technique called orthogonal FDMA (OFDMA). It motivates us to study (O)FDMA-based multi-user IRS communications to clarify which user scheduling and passive beamforming are favorable under this non-frequency-selective reflection environment. Theoretical analysis and numerical evaluation reveal that (O)FDMA does not need user scheduling when there are a few users. If the number of users becomes large, neither user scheduling nor IRS reflection optimization is necessary. These findings help substantially simplify the design of (O)FDMA-based IRS communications.
\end{abstract}

%
\IEEEpeerreviewmaketitle

\section{Introduction}
Recently, intelligent reflecting surface (IRS), a.k.a. reconfigurable intelligent surface, has drawn much attention from academia and industry because of its potential of reaping high performance in a green and cost-effective way \cite{Ref_liu2021reconfigurable}.
Particularly, IRS is a planar meta-surface consisting of a large number of passive, lightweight, and cheap reflection elements.  A smart controller dynamically adjusts the phase shift of each element, thereby collaboratively achieving smart propagation environment for signal amplification or interference suppression \cite{Ref_wu2019intelligent}. Conventional technologies, such as \textit{dense and heterogeneous network deployment}, \textit{massive antennas}, and \textit{ultra-wide bandwidth}, can improve coverage and capacity effectively but incur high capital and operational expenditures, much energy consumption, and severe network interference. In contrast, IRS not only reflects signals in a full-duplex and noise-free way \cite{Ref_renzo2020reconfigurable} but also lowers hardware and energy costs substantially thanks to the use of passive components. Consequently, it is widely recognized that IRS can serve as a key technological enabler for the upcoming sixth-generation (6G) wireless system to meet more stringent performance requirements than its predecessor \cite{Ref_jiang2021road}.

Previous studies have investigated different aspects for IRS-aided wireless communications, e.g., passive beamforming \cite{Ref_wu2019intelligent}, reflected channel estimation \cite{Ref_wang2020channel}, hardware constraints \cite{Ref_zhi2021uplink}, and the interplay with other technologies, such as orthogonal frequency-division multiplexing (OFDM) \cite{Ref_zheng2020intelligent}, multi-input multi-output (MIMO) \cite{Ref_hu2018beyond}, hybrid beamforming \cite{Ref_di2020hybrid}, and Terahertz communications \cite{Ref_jiang2022dualbeam}.
The majority of prior works focus on point-to-point IRS-aided communications, which merely considers a single user for simplicity. Nevertheless, a practical system needs to accommodate many users simultaneously, especially in scenarios such as massive connectivity, raising the problem of multiple access. Usually, the phase shift of an element is adaptively tuned across different time instants (i.e., time-selective). Due to the hardware constraint, it is not frequency-selective, namely the phase shift cannot be different across frequency subchannels. Hence, the integration of IRS imposes a fundamental challenge to frequency-multiplexing approaches such as frequency-division multiple access (FDMA), including the widely adopted technique called orthogonal FDMA (OFDMA) \cite{Ref_jiang2016ofdm}.

In \cite{Ref_zheng2020intelligent_COML}, the authors revealed that non-orthogonal multiple access (NOMA) and time-division multiple access (TDMA) are superior to FDMA in terms of spectral efficiency. However, NOMA needs complex signal processing for successive interference cancellation (SIC) \cite{Ref_chen2019toward}, while TDMA requires stringent time synchronization among users.
It is envisioned that FDMA can still play an important role in some particular 6G scenarios such as IRS-assisted Internet of Things \cite{Ref_chu2022resource} due to its simple implementation. Meanwhile, intelligent surfaces can be transparently installed in legacy FDMA-based networks for performance enhancement, especially for OFDMA-empowered 4G LTE and 5G NR. Hence, it is worth making clear the behaviours of (O)FDMA-based multi-user IRS systems in the absence of frequency-selective reflection, which is not reported yet in the literature to the best knowledge of the authors. For simplicity, we use FDMA hereinafter to represent both legacy FDMA and OFDMA techniques.

Since the IRS in FDMA-based systems can only be optimized for a particular user/subchannel, it raises two fundamental questions: i) which user should be selected for optimizing passive beamforming? and ii) how to optimize the passive beamforming?
In this paper, therefore, we investigate the effects of multi-user scheduling and reflection optimization in an IRS-aided FDMA system (interchangeably using the term \textit{FDMA-IRS} hereinafter). Spectral efficiency for different user-scheduling and passive-beamforming strategies are analytically derived, while a simulation setup consisting of a cell-edge area and a cell-center area is deliberately designed to compare their performance with TDMA and power-domain NOMA.  Theoretical analysis and numerical evaluation reveal a meaningful outcome. That is, FDMA-IRS does not need user scheduling when there are a few users. If the number of users becomes large, \textit{neither} user scheduling \textit{nor} passive-beamforming optimization is necessary. This findings helps substantially simplifying the design of FDMA-IRS, where at least the estimation of cascaded channels, the IRS tuning, the smart controller, and its backhaul are not needed any more.

We organize the rest of this paper as follows: Section II gives the system model. Section III analyzes user scheduling and reflection optimization for FDMA-based IRS communications. Simulation setup and some examples of numerical results are demonstrated in Section IV. Finally, Section V summarizes this paper.

\section{System Model}
\begin{figure}[!tbph]
    \centering
    \includegraphics[width=0.42\textwidth]{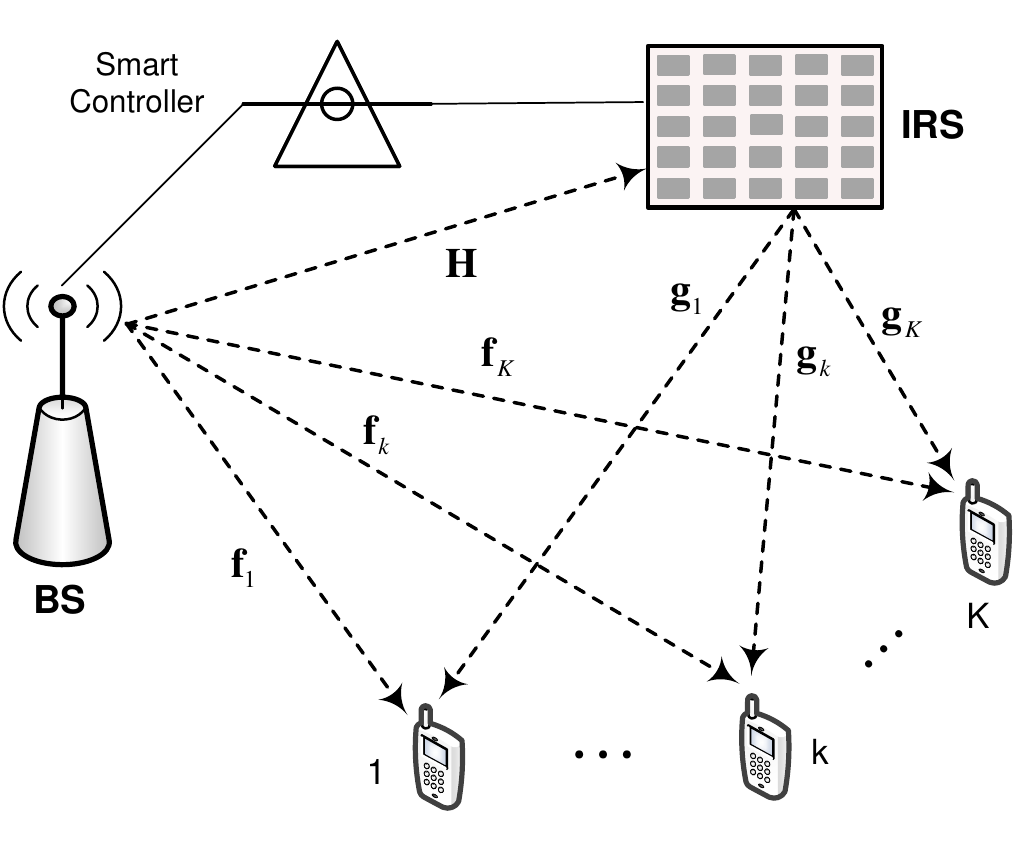}
    \caption{Schematic diagram of an IRS-aided multi-user MIMO system, consisting of a multi-antenna BS, $K$ single-antenna UE, and a reflecting surface with $N$ IRS elements.  }
    \label{diagram:system}
\end{figure}

As illustrated in \figurename \ref{diagram:system}, this paper focuses on the downlink of an IRS-assisted multi-user MIMO  system, where an intelligent surface with $N$ reflecting elements is deployed to aid the transmission from an $N_b$-antenna BS to $K$ single-antenna user equipment (UE). We write
\begin{align}
    \mathbf{f}_{k}=\Bigl[f_{k1},f_{k2},\ldots,f_{kN_b}\Bigr]^T
\end{align}
to denote the $N_b\times 1$ channel vector between the BS and the $k^{th}$ UE, and
\begin{align}
    \mathbf{g}_{k}=\Bigl[g_{k1},g_{k2},\ldots,g_{kN}\Bigr]^T
\end{align}
to denote the $N\times 1$ channel vector between the IRS and UE $k$. Denoting  the channel vector from the BS to the $n^{th}$ reflecting element by $\mathbf{h}_{n}=[h_{n1},h_{n2},\ldots,h_{nN_b}]^T$, the channel matrix from the BS to the IRS is expressed as $\mathbf{H}\in \mathbb{C}^{N\times N_b}$, where the $n^{th}$ row of $\mathbf{H}$ equals to $\mathbf{h}_n^T$.
Since the line-of-sight (LOS) paths from either the BS or the IRS to UEs may be blocked, the corresponding small-scale fading follows Rayleigh distribution. Consequently, $f$ and $g$ are circularly symmetric complex Gaussian random variables denoted by $f\sim \mathcal{CN}(0,\sigma_f^2)$ and $g\sim \mathcal{CN}(0,\sigma_g^2)$, respectively.
In contrast, a favourable location  is deliberately selected for the IRS to exploit an LOS path to the fixed BS without any blockage, resulting in Rician fading, i.e.,
\begin{equation}\label{EQNIRQ_LSFadingdirect}
    \mathbf{H}=\sqrt{\frac{\Gamma\sigma_h^2}{\Gamma+1}}\mathbf{H}_{LOS} + \sqrt{\frac{\sigma_h^2}{\Gamma+1}}\mathbf{H}_{NLOS}
\end{equation}
with the Rician factor $\Gamma$, the LOS component $\mathbf{H}_{LOS}$,  the multipath component $\mathbf{H}_{NLOS}$ consisting of independent entries that follow $\mathcal{CN}(0,1)$, and the BS-IRS path loss $\sigma_h^2$.

Since the IRS is a passive device, time-division duplexing (TDD) operation is usually adopted to simplify channel estimation. The users send pilot signals in the uplink training and the BS estimates the uplink instantaneous channel state information (CSI), which is used for optimizing downlink data transmission due to channel reciprocity.  To characterize the theoretical performance, the analysis is conducted under the assumption that the BS perfectly knows the  CSI of all involved channels, as most prior works \cite{Ref_wu2019intelligent, Ref_renzo2020reconfigurable, Ref_renzo2020smart}.
A smart controller adaptively adjusts the phase shift of each reflecting element based on the instantaneous CSI acquired through periodic channel estimation \cite{Ref_wang2020channel}. The reflection of a typical element $n$ is mathematically modeled by a coefficient $\epsilon_{n}=a_{n} e^{j\phi_{n}}$, where $\phi_{n}\in [0,2\pi)$ denotes an induced phase shift, and $a_{n}\in [0,1]$ stands for amplitude attenuation. As mentioned by \cite{Ref_wu2019intelligent}, $a_{n}=1$, $\forall n$ is the optimal attenuation that maximizes the strength of the received signal and simplifies the implementation complexity. Hence, the reflection optimization only focuses on the phase shifts $\phi_{n}$, $\forall n$.  By ignoring hardware impairments such as quantified phase shifts \cite{Ref_wu2020beamforming} and phase noise \cite{Ref_jiang2022impact}, the $k^{th}$ UE observes the received signal
\begin{equation} \label{eqn_systemModel}
    r_k=\sqrt{P_t}\Biggl( \sum_{n=1}^N g_{kn} e^{j\phi_{n}} \mathbf{h}_{n}^T + \mathbf{f}_{k}^T \Biggr) \mathbf{s} + n_k,
\end{equation}
where $\mathbf{s}$ denotes the vector of transmitted signals over the BS antenna array, $P_t$ expresses the transmit power, $n_k$ is additive white Gaussian noise (AWGN) with zero mean and variance $\sigma_n^2$, namely $n_k\sim \mathcal{CN}(0,\sigma_n^2)$. Define
\begin{equation}
    \boldsymbol{\Theta}_0=\mathrm{diag}\Bigl\{e^{j\phi_{1}},e^{j\phi_{2}},\ldots,e^{j\phi_{N}}\Bigr\},
\end{equation} \eqref{eqn_systemModel} can be rewritten in matrix form as
\begin{equation} \label{EQN_IRS_RxSignal_Matrix}
    r_k= \sqrt{P_t}\Bigl(\mathbf{g}_k^T \boldsymbol{\Theta}_0 \mathbf{H} +\mathbf{f}_k^T\Bigr)\mathbf{s} +n_k.
\end{equation}

\section{User Scheduling and Optimization Design}
The fundamental of FDMA lies on the division of the whole system bandwidth into $K$ orthogonal subchannels. Each user occupies a dedicated subchannel over the entire time. At the $k^{th}$ subchannel, the BS antenna array applies linear beamforming $\mathbf{w}_k\in \mathbb{C}^{N_b\times 1}$, where $\|\mathbf{w}_k\|^2\leqslant 1$, to send the information symbol $s_k$ intended for user $k$, which satisfies $\mathbb{E}\left[|s_k|^2\right]=1$.  For simplicity, the transmit power of the BS is equally allocated among subchannels, i.e., $P_t=P_d/K$, where $P_d$ expresses the power constraint of the BS.

Substituting $\mathbf{s}=\mathbf{w}_k s_k$ and $P_t=P_d/K$ into \eqref{EQN_IRS_RxSignal_Matrix}, we obtain
\begin{equation}
    r_k= \sqrt{\frac{P_d}{K}}\Bigl(\mathbf{g}_k^T \boldsymbol{\Theta}_0 \mathbf{H} +\mathbf{f}_k^T\Bigr)\mathbf{w}_k s_k +n_k.
\end{equation}
Thus, the achievable spectral efficiency of user $k$ is formulated as
\begin{equation}
    R_k=\frac{1}{K}\log\left(1+\frac{\frac{P_d}{K} \Bigl|\bigl(\mathbf{g}_k^T \boldsymbol{\Theta}_0 \mathbf{H} +\mathbf{f}_k^T\bigr)\mathbf{w}_k\Bigr|^2 }{\frac{\sigma_n^2}{K}} \right),
\end{equation}
where the factor $1/K$ is applied since the signal transmission of each user occupies only $1/K$ of the total bandwidth.
As a result, the sum rate of the IRS-aided FDMA system is
 \begin{equation} \label{EQNIRQ_sumRate}
     C= \sum_{k=1}^K \frac{1}{K}\log\left(1+\frac{P_d \Bigl|\bigl(\mathbf{g}_{k}^T \boldsymbol{\Theta}_0 \mathbf{H} +\mathbf{f}_{k}^T\bigr)\mathbf{w}_{k}\Bigr|^2 }{\sigma_n^2} \right).
  \end{equation}
\subsection{Multi-User Scheduling}
Unlike an IRS-aided TDMA system, where the set of phase shifts at each slot can be dynamically adjusted for its assigned user,  the IRS in an FDMA system can be optimized only for a particular user $\hat{k}$ by setting $\boldsymbol{\Theta}_{\hat{k}}$. The remaining users $k\neq \hat{k}$ over other subchannels have to utilize this common configuration with $\boldsymbol{\Theta}_{\hat{k}}$ and suffer from phase-unaligned reflection. That is because the hardware constraint of IRS elements, which can be tuned in \textit{time-selective} but not \textit{frequency-selective}.
Hence, it raises a particular issue of \textit{user scheduling} in FDMA: which user should be selected for passive beamforming optimization?  To get a comprehensive view, we design the following strategies:
\begin{itemize}
    \item \textit{Exhaustive Search:}  If the system applies the IRS to aid the signal transmission of a typical user $k$, the set of phase shifts is adjusted according to the CSI of this user, i.e., $\boldsymbol{\Theta}_0=f(\mathbf{f}_k, \mathbf{g}_k, \mathbf{H} )$, where $f(\cdot)$ denotes a certain optimization function. In this case, the achievable sum rate denoted by $C(k)$ can be computed by \eqref{EQNIRQ_sumRate}. The best user selection, which achieves the maximal sum rate, is determined by exhaustively setting each user as the IRS-aided user   \begin{equation}
        \hat{k}=\arg\max_{k\in \mathcal{K}} C(k),
    \end{equation} where $\mathcal{K}=\{1,2,\ldots,K\}$ represents the set of users.
    \item \textit{The Nearest User:} The philosophy behind is that the system performance might be improved by applying the IRS to aid the user that has the smallest distance (equivalent to the largest large-scale fading or average received power) to the IRS. Without losing generality, the user is selected according to
    \begin{equation}
        \hat{k}=\arg\max_{k\in \mathcal{K}} \mathbb{E}\left[ \|\mathbf{g}_k\|^2 \right],
    \end{equation}
    where $\| \cdot\|$ expresses the Frobenius norm, and $\mathbb{E}$ denotes the statistical expectation.
    \item \textit{The Farthest User:} For comparison, we can also optimize the phase shifts for the user that has the largest distance to the IRS, i.e.,
    \begin{equation}
        \hat{k}=\arg\min_{k\in \mathcal{K}} \mathbb{E}\left[ \|\mathbf{g}_k\|^2 \right].
    \end{equation}
    \item \textit{Random Selection} where the IRS adjusts its phase shifts according to the CSI of a user that is randomly selected from $\mathcal{K}$. In other words,  the IRS is optimized as $\boldsymbol{\Theta}_0=f(\mathbf{f}_k, \mathbf{g}_k, \mathbf{H} )$ given a random user $k$. Note that random user selection differs from random IRS reflection, where the phase shifts are random without optimization, independent of the CSI of any user.
\end{itemize}

\subsection{Joint Reflection Optimization}
When the best user $\hat{k}$ is determined through user scheduling, the system needs the joint optimization of active beamforming at the BS and passive reflection at the IRS to aid the signal transmission of $\hat{k}$. Temporarily ignoring other users $k\neq \hat{k}$, the closed-form solutions for optimal active beamforming and optimal reflection can be obtained through an alternating method as presented below.
To maximize the instantaneous signal-to-noise ratio (SNR) of user $\hat{k}$, i.e.,
\begin{equation} \label{IRS_EQN_spectralEfficiency}
    \gamma_{\hat{k}}=\frac{P_d \Bigl|\bigl(\mathbf{g}_{\hat{k}}^T \boldsymbol{\Theta}_0 \mathbf{H} +\mathbf{f}_{\hat{k}}^T\bigr)\mathbf{w}_{\hat{k}}\Bigr|^2 }{\sigma_n^2},
\end{equation}
we formulate the following optimization problem
\begin{equation}
\begin{aligned} \label{eqnIRS:optimizationMRTvector}
\max_{\boldsymbol{\Theta}_0,\:\mathbf{w}_{\hat{k}}}\quad &  \biggl|\Bigl(\mathbf{g}_{\hat{k}}^T \boldsymbol{\Theta}_0 \mathbf{H} +\mathbf{f}_{\hat{k}}^T\Bigr)\mathbf{w}_{\hat{k}}\biggr|^2\\
\textrm{s.t.} \quad & \|\mathbf{w}_{\hat{k}}\|^2\leqslant 1\\
  \quad & \phi_{n}\in [0,2\pi), \: \forall n=1,2,\ldots,N,
\end{aligned}
\end{equation}
which is non-convex because their objective function is not jointly concave with respect to $\boldsymbol{\Theta}_0$ and $\mathbf{w}_{\hat{k}}$. To solve this problem, we can alternately optimize $\boldsymbol{\Theta}_0$ and $\mathbf{w}_{\hat{k}}$ in an iterative manner \cite{Ref_wu2019intelligent}.

Without loss of generality, the maximal-ratio transmission (MRT) for the direct link can be applied as the initial value of the transmit vector, i.e., $\mathbf{w}_{\hat{k}}^{(0)}=\mathbf{f}_{\hat{k}}^*/\|\mathbf{f}_{\hat{k}}\|$.
Thus, \eqref{eqnIRS:optimizationMRTvector} is simplified to
\begin{equation}  \label{eqnIRS:optimAO}
\begin{aligned} \max_{\boldsymbol{\Theta}_0}\quad &  \biggl|\Bigl(\mathbf{g}_{\hat{k}}^T \boldsymbol{\Theta}_0 \mathbf{H} +\mathbf{f}_{\hat{k}}^T\Bigr)\mathbf{w}_{\hat{k}}^{(0)}\biggr|^2\\
\textrm{s.t.}  \quad & \phi_{n}\in [0,2\pi), \: \forall n=1,2,\ldots,N.
\end{aligned}
\end{equation}
The objective function is still non-convex but it enables a closed-form solution through applying the well-known triangle inequality
\begin{equation}
    \biggl|\Bigl(\mathbf{g}_{\hat{k}}^T \boldsymbol{\Theta}_0 \mathbf{H} +\mathbf{f}_{\hat{k}}^T\Bigr)\mathbf{w}_{\hat{k}}^{(0)}\biggr| \leqslant \biggl|\mathbf{g}_{\hat{k}}^T \boldsymbol{\Theta}_0 \mathbf{H} \mathbf{w}_{\hat{k}}^{(0)}\biggr| +\biggl|\mathbf{f}_{\hat{k}}^T\mathbf{w}_{\hat{k}}^{(0)}\biggr|.
\end{equation}
The equality achieves if and only if
\begin{equation}
    \arg\left (\mathbf{g}_{\hat{k}}^T \boldsymbol{\Theta}_0 \mathbf{H} \mathbf{w}_{\hat{k}}^{(0)}\right)= \arg\left(\mathbf{f}_{\hat{k}}^T\mathbf{w}_{\hat{k}}^{(0)}\right)\triangleq \varphi_{0},
\end{equation}
where $\arg(\cdot)$ stands for the component-wise phase of a complex vector.

Define $\mathbf{b}_0=\left[b_{1},b_{2},\ldots,b_{N}\right]^H$ with $b_{n}=e^{j\phi_{n}}$ and  $\boldsymbol{\chi}_{\hat{k}}=\mathrm{diag}(\mathbf{g}_{\hat{k}}^T)\mathbf{H}\mathbf{w}_{\hat{k}}^{(0)}\in \mathbb{C}^{N\times 1}$, we have $\mathbf{g}_{\hat{k}}^T \boldsymbol{\Theta}_0 \mathbf{H} \mathbf{w}_{\hat{k}}^{(0)}=\mathbf{b}_0^H\boldsymbol{\chi}_{\hat{k}}\in \mathbb{C} $.
Ignore the constant term $\bigl|\mathbf{f}_{\hat{k}}^T\mathbf{w}_{\hat{k}}^{(0)}\bigr|$, \eqref{eqnIRS:optimAO} is transformed to
\begin{equation}  \label{eqnIRS:optimizationQ}
\begin{aligned} \max_{\boldsymbol{\mathbf{b}_0}}\quad &  \Bigl|\mathbf{b}_0^H\boldsymbol{\chi}_{\hat{k}}\Bigl|\\
\textrm{s.t.}  \quad & |b_{n}|=1, \: \forall n=1,2,\ldots,N,\\
  \quad & \arg(\mathbf{b}_0^H\boldsymbol{\chi}_{\hat{k}})=\varphi_{0}.
\end{aligned}
\end{equation}
The solution for \eqref{eqnIRS:optimizationQ} can be derived as
\begin{equation} \label{eqnIRScomplexityQ}
    \mathbf{b}_0^{(1)}=e^{j\left(\varphi_{0}-\arg(\boldsymbol{\chi}_{\hat{k}})\right)}=e^{j\left(\varphi_{0}-\arg\left( \mathrm{diag}(\mathbf{g}_{\hat{k}}^T)\mathbf{H}\mathbf{w}_{\hat{k}}^{(0)}\right)\right)}.
\end{equation}
Equivalently,
\begin{align} \nonumber \label{IRSeqnOptimalShift}
    \phi_{n}^{(1)}&=\varphi_{0}-\arg\left(g_{\hat{k}n}\mathbf{h}_{n}^T\mathbf{w}_{\hat{k}}^{(0)}\right)\\&=\varphi_{0}-\arg\left(g_{\hat{k}n}\right)-\arg\left(\mathbf{h}_{n}^T\mathbf{w}_{\hat{k}}^{(0)}\right),
\end{align}
where $\mathbf{h}_{n}^T\mathbf{w}_{\hat{k}}^{(0)}\in \mathbb{C} $ can be regarded as an effective SISO channel perceived by the $n^{th}$ reflecting element, combining the effects of transmit beamforming $ \mathbf{w}_{\hat{k}}^{(0)}$ and channel response $\mathbf{h}_{n}$. In this regard, \eqref{IRSeqnOptimalShift} implies that an IRS reflector should be tuned such that the residual phase of each reflected signal is aligned with the phase of the signal over the direct link for coherent combining at the receiver.

Once the phase shifts at the first iteration, i.e., $    \boldsymbol{\Theta}^{(1)}_0=\mathrm{diag}\left\{e^{j\phi_{1}^{(1)}},e^{j\phi_{2}^{(1)}},\ldots,e^{j\phi_{N}^{(1)}}\right\}$ are determined, the optimization is alternated to update $\mathbf{w}_{\hat{k}}$. The BS applies MRT a.k.a. matched filtering to maximize the strength of the desired signal, resulting in
\begin{equation}  \label{EQN_IRS_TXBF}
    \mathbf{w}_{\hat{k}}^{(1)} = \frac{\Bigl(\mathbf{g}_{\hat{k}}^T \boldsymbol{\Theta}^{(1)}_0 \mathbf{H} +\mathbf{f}_{\hat{k}}^T\Bigr)^H}{\Bigl\|\mathbf{g}_{\hat{k}}^T \boldsymbol{\Theta}_0^{(1)} \mathbf{H} +\mathbf{f}_{\hat{k}}^T\Bigr\|}.
\end{equation}
After the completion of the first iteration, the BS gets $\boldsymbol{\Theta}^{(1)}_0$ and $\mathbf{w}_{\hat{k}}^{(1)}$, which serve as the initial input for the second iteration to derive $\boldsymbol{\Theta}^{(2)}_0$ and $\mathbf{w}_{\hat{k}}^{(2)}$.
This process iterates until the convergence is achieved with the optimal beamformer $\mathbf{w}_{\hat{k}}^{\star}$ and  optimal reflection $\boldsymbol{\Theta}_0^{\star}$. Here, the convergence means the objective value of \eqref{eqnIRS:optimizationMRTvector} is non-increasing over iterations, which can be implemented flexibly.

Applying $\mathbf{w}_{\hat{k}}^{\star}$ and  $\boldsymbol{\Theta}_0^{\star}$ into \eqref{IRS_EQN_spectralEfficiency}, we derive the achievable spectral efficiency of user $\hat{k}$ as
\begin{equation}
    R_{\hat{k}}=\frac{1}{K}\log\left(1+\frac{P_d \Bigl|\bigl(\mathbf{g}_{\hat{k}}^T \boldsymbol{\Theta}_0^\star \mathbf{H} +\mathbf{f}_{\hat{k}}^T\bigr)\mathbf{w}_{\hat{k}}^\star\Bigr|^2 }{\sigma_n^2} \right).
\end{equation}
Given the reflection $\boldsymbol{\Theta}^{\star}_{0}$,  what the remaining $K-1$ users can do is to achieve partial optimization (in contrast to the joint optimization for user $\hat{k}$) by updating their respective active beamforming. The MRT beamformer for user $k \neq \hat{k}$, $\forall k=1,2,\ldots,K$ is optimized as
\begin{equation}  \label{EQN_IRS_matchedFilter}
    \mathbf{w}_{k}^{\star} = \frac{\Bigl(\mathbf{g}_{k}^T \boldsymbol{\Theta}^{\star}_{0} \mathbf{H} +\mathbf{f}_{k}^T\Bigr)^H}{\Bigl\|\mathbf{g}_{k}^T \boldsymbol{\Theta}_{0}^{\star} \mathbf{H} +\mathbf{f}_{k}^T\Bigr\|}.
\end{equation}
Substituting the optimal parameters into \eqref{EQNIRQ_sumRate}, the sum rate of the FDMA IRS system is obtained as
 \begin{align}
     C&= \underbrace{\frac{1}{K}\log\left(1+\frac{P_d \Bigl|\bigl(\mathbf{g}_{\hat{k}}^T \boldsymbol{\Theta}_{0}^\star \mathbf{H} +\mathbf{f}_{\hat{k}}^T\bigr)\mathbf{w}_{\hat{k}}^\star\Bigr|^2 }{\sigma_n^2} \right)}_{\text{Joint-Optimized\:User}}  \\ \nonumber
     &+\underbrace{\sum_{k=1,k\neq \hat{k}}^K \frac{1}{K}\log\left(1+\frac{P_d \Bigl|\bigl(\mathbf{g}_{k}^T \boldsymbol{\Theta}_{0}^\star \mathbf{H} +\mathbf{f}_{k}^T\bigr)\mathbf{w}_{k}^\star\Bigr|^2 }{\sigma_n^2} \right)}_{\text{Partial-Optimized\:Users}}\\ \nonumber
     &= \frac{1}{K}\log\left(1+\Bigl\| \mathbf{g}_{\hat{k}}^T \boldsymbol{\Theta}_{0}^\star \mathbf{H} +\mathbf{f}_{\hat{k}}^T \Bigr\|^2\frac{P_d }{\sigma_n^2} \right)  \\ \nonumber
     &+\sum_{k=1,k\neq \hat{k}}^K \frac{1}{K}\log\left(1+ \bigl\|\mathbf{g}_{k}^T \boldsymbol{\Theta}_{0}^\star \mathbf{H} +\mathbf{f}_{k}^T \bigr\|^2 \frac{P_d }{\sigma_z^2} \right).
  \end{align}

\subsection{Random Reflection}
In addition to the joint optimization of passive reflection and active beamforming, we study the simplest solution as a baseline. Randomly setting the phase shifts of IRS elements denoted by $\Theta_r$, each entry of which takes value uniformly and randomly from $[0,2\pi)$. Then, each subchannel just optimizes its transmit beamforming as
\begin{equation}
    \mathbf{w}_{k}^{\star} = \frac{\Bigl(\mathbf{g}_{k}^T \boldsymbol{\Theta}_{r} \mathbf{H} +\mathbf{f}_{k}^T\Bigr)^H}{\Bigl\|\mathbf{g}_{k}^T \boldsymbol{\Theta}_{r} \mathbf{H} +\mathbf{f}_{k}^T\Bigr\|}.
\end{equation}
Since the phase shifts are random without the knowledge of CSI, the channel estimation of cascaded channels is avoided, substantially simplifying the system complexity.

\begin{figure}[!ht]
    \centering
    \includegraphics[width=0.42\textwidth]{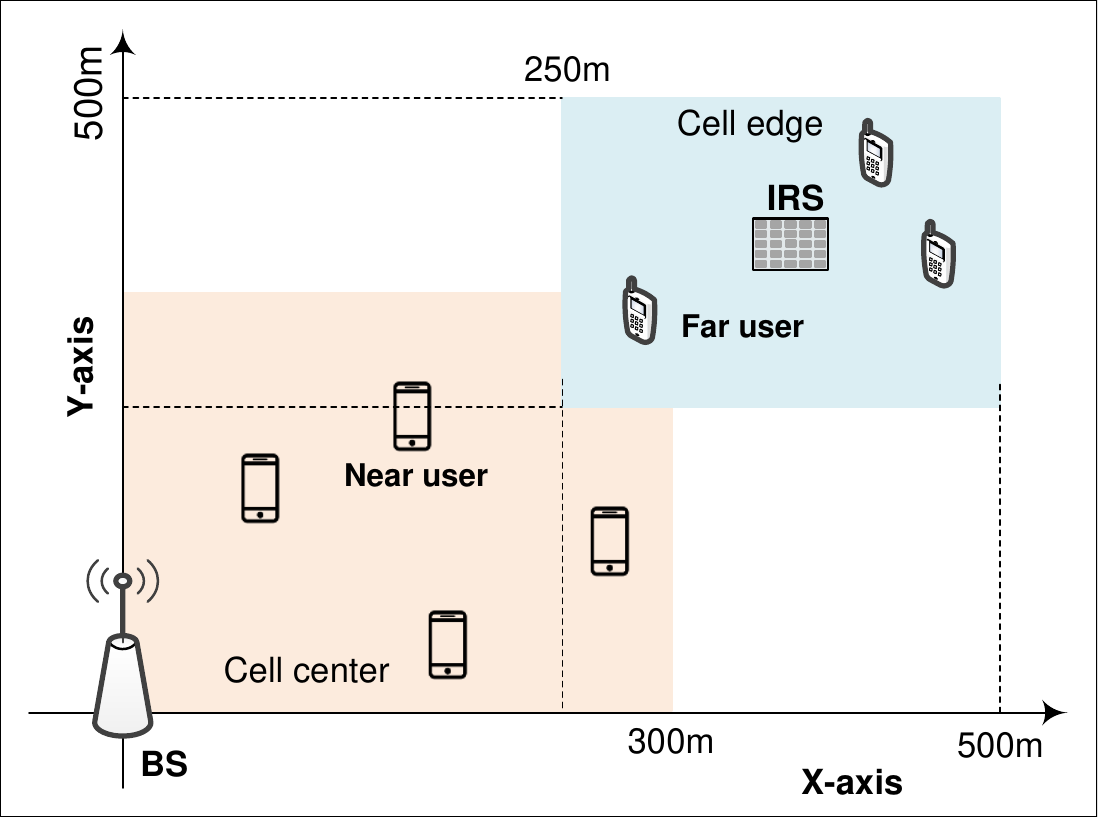}
    \caption{Simulation scenario of a multi-user IRS system, where the cell coverage is comprised of a cell-center area and a cell-edge area. }
    \label{diagram:Simulation}
\end{figure}

\section{Performance Evaluation}
\begin{figure*}[!t]
\centerline{
\subfloat[]{
\includegraphics[width=0.43\textwidth]{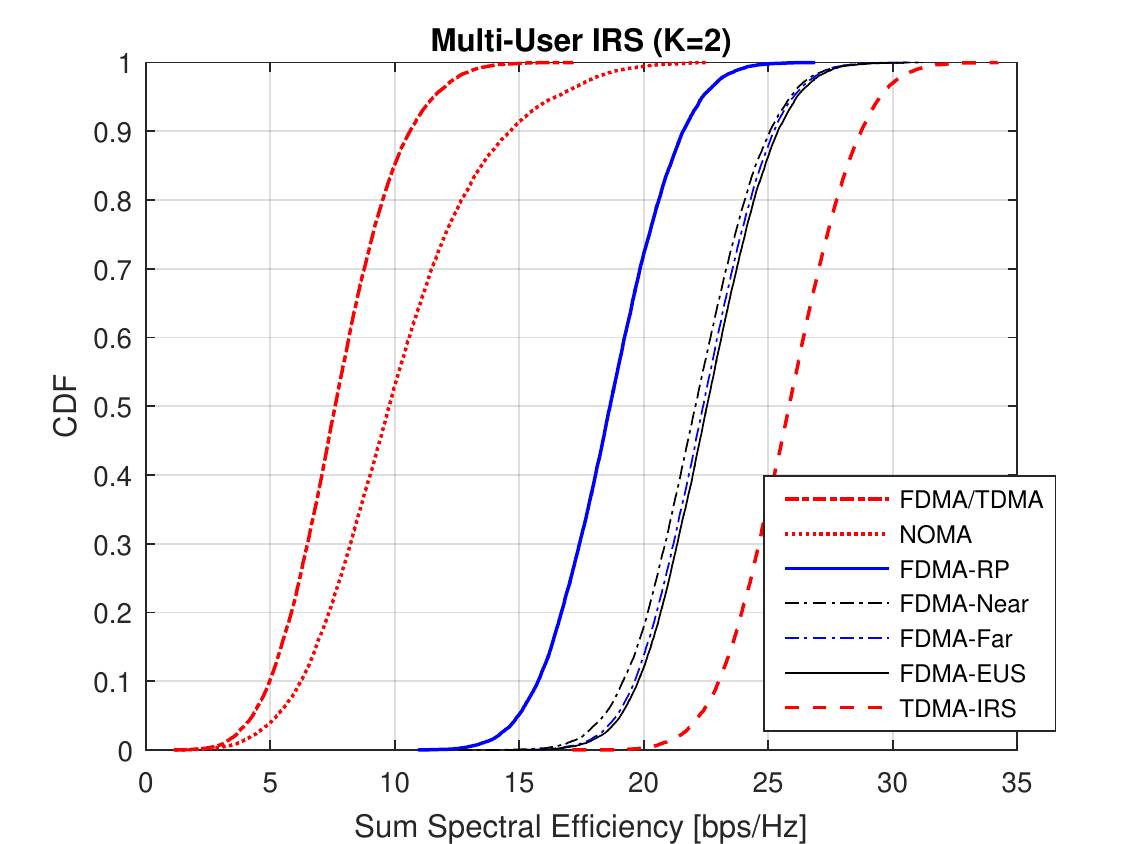}
\label{Fig_results1}
}
\hspace{10mm}
\subfloat[]{
\includegraphics[width=0.43\textwidth]{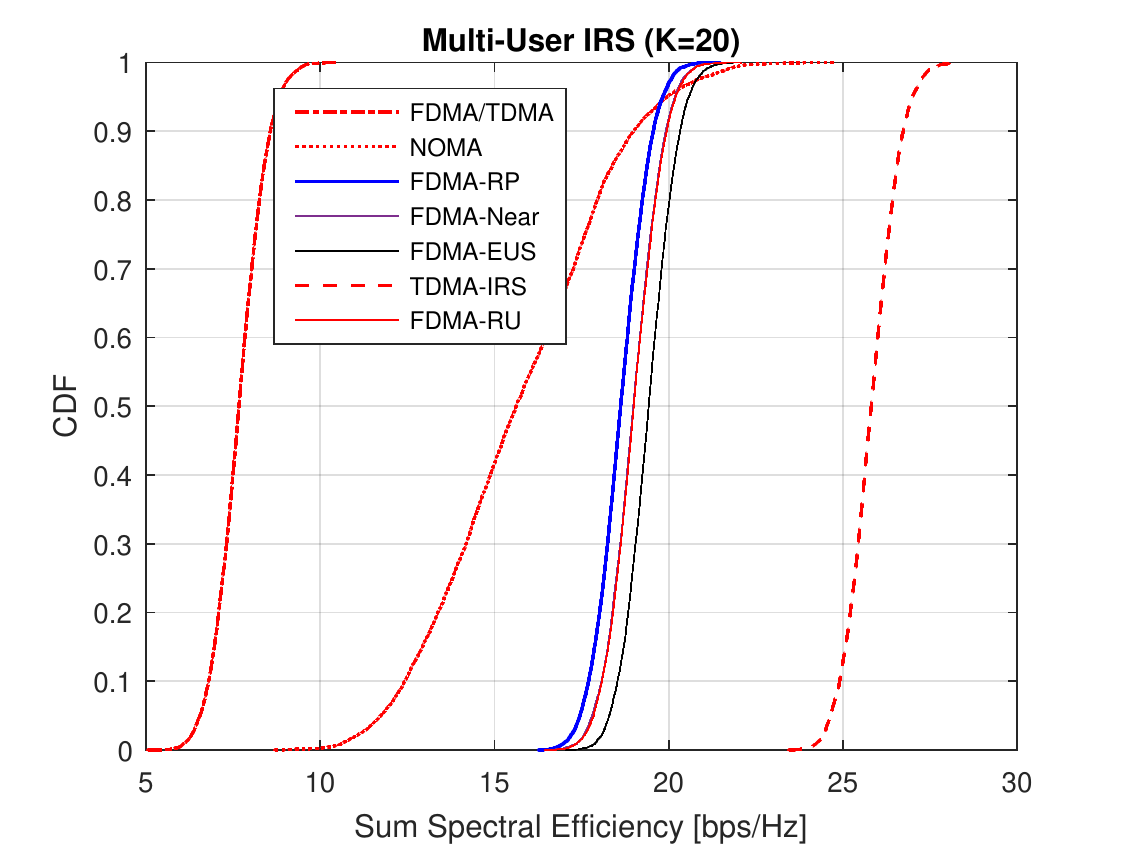}
\label{Fig_results2}
}}
\caption{Performance comparisons of TDMA, NOMA and FDMA with different user scheduling and reflection optimization schemes in an IRS-aided multi-user MIMO systems: (a) CDFs of the sum spectral efficiency with two users, and (b) CDFs of the sum spectral efficiency with twenty users. }
\label{Fig_Result}
\end{figure*}

In this section, we illustrate some numerical results of achievable spectral efficiency to compare the performance of FDMA, TDMA and NOMA. Different user scheduling and reflection optimization strategies for FDMA are evaluated.
As shown in \figurename \ref{diagram:Simulation}, we consider the cell coverage consisting of a cell-center area and a cell-edge area. The BS equipped with $N_b=16$ antennas is located at the original point $(0,0)$ of the coordinate system, while a surface containing $N=200$ reflecting elements is installed at the center of the cell-edge area, with the coordinate $(375\si{\meter},375\si{\meter})$.  Half of the users are far users that distribute randomly over the cell-edge area, while the other half of users are near users that distribute randomly over the cell-edge area. The maximum transmit power of BS is $P_d=20\mathrm{W}$ over a signal bandwidth of $B_w=20\mathrm{MHz}$, referring to the practical 3GPP LTE specification.  We apply the formula $\sigma_n^2=\kappa\cdot B_w\cdot T_0\cdot N_f$ to compute the variance of AWGN, where $\kappa$ is the Boltzmann constant, the temperature $T_0=290 \mathrm{Kelvin}$, and the noise figure $N_f=9\mathrm{dB}$.

The  large-scale fading is computed by $\sigma^2=10^\frac{\mathcal{P}+\mathcal{S}}{10}$, where $\mathcal{P}$ is distance-dependent path loss, $\mathcal{S}$ stands for \textit{log-normal} shadowing denoted by $\mathcal{S}\sim \mathcal{N}(0,\sigma_{sd}^2)$, and the standard derivation is generally set to $\sigma_{sd}=8\mathrm{dB}$. As  \cite{Ref_jiang2021cellfree}, this paper employs the COST-Hata model to calculate $\mathcal{P}$, i.e.,
\begin{equation} \label{eqn:CostHataModel}
    \mathcal{P}=
\begin{cases}
-\mathcal{P}_0-35\lg(d), &  d>d_1 \\
-\mathcal{P}_0-15\lg(d_1)-20\lg(d), &  d_0<d\leq d_1 \\
-\mathcal{P}_0-15\lg(d_1)-20\lg(d_0), &  d\leq d_0
\end{cases},
\end{equation}
where $d$ represents the propagation distance,  $d_0$ and $d_1$ are reference points, which take values $d_0=10\mathrm{m}$ and $d_1=50\mathrm{m}$, respectively. The first term is formulated as  \begin{IEEEeqnarray}{ll}
\mathcal{P}_0=46.3&+33.9\lg\left(f_0\right)-13.82\lg\left(h_{S}\right)\\ \nonumber
 &-\Bigl[1.1\lg(f_0)-0.7\Bigr]h_{T}+1.56\lg\left(f_0\right)-0.8,
\end{IEEEeqnarray} which equals $\mathcal{P}_0=140.72\mathrm{dB}$ when the carrier frequency $f_0=1.9\mathrm{GHz}$, the antenna height of BS or IRS $h_{S}=15\mathrm{m}$, and the UE antenna height $h_{T}=1.65\mathrm{m}$.
In contrast, the path loss for the LOS channel between the BS and IRS can be computed through
\begin{equation}
    L(d)=\frac{L_0}{d^{-\alpha}},
\end{equation}
where $L_0=\SI{-30}{\decibel}$ is the path loss measured at the reference distance of \SI{1}{\meter}, $\alpha=2$ means the free-space path loss exponent, and the Rician factor is set to $K=5$.

For a comprehensive view, our evaluation compares different schemes and configurations as many as possible, including: 1) TDMA without the aid of IRS, where a radio frame is orthogonally divided into $K$ time slots. At the $k^{th}$ slot, the BS sends the information symbol $s_k$ to user $k$. The MRT $\mathbf{w}_k^\star=\mathbf{f}_k^*/\|\mathbf{f}_k\|$ is applied to achieve matched filtering in terms of the BS-UE direct link.   2) NOMA: the transmitter superimposes all information symbols into a single waveform as $\mathbf{s}=\sum_{k=1}^K \sqrt{\alpha_k } \mathbf{w}_{k} s_k$,
where  $\alpha_k$ represents the power allocation coefficient subjecting to $\sum_{k=1}^K\alpha_k\leqslant 1$. The optimal order of successive interference cancellation is detecting the user with the weakest channel gain first. Then, the corresponding component due to the weakest user is subtracted from the received signal at each user. In the second iteration, each user decodes the second weakest user using the residual signal. The cancellation iterates until each user gets its own symbol. 3) FDMA without the aid of IRS, where the BS applies the MRT $\mathbf{w}_k^\star=\mathbf{f}_k^*/\|\mathbf{f}_k\|$ to send the information symbol $s_k$ at the $k^{th}$ subchannel.  4) TDMA in the IRS-aided system denoted by TDMA-IRS, where the IRS applies time-selective phase shifts at each slot. At the $k^{th}$ slot, alternating optimization is applied to jointly optimize active beamforming and passive reflection. 5) FDMA-RP means \textit{randomly} setting the \textit{phase} shifts of the surface  in the IRS-aided FDMA system, and then each subchannel optimizes its own transmit vector $\mathbf{w}_k$ given the random but known reflection situation $\boldsymbol{\Theta}_r$. 6) FDMA-Near means using the CSI of the nearest user of the IRS to optimize the phase shifts, and then each subchannel optimizes its own transmit vector $\mathbf{w}_k$ in terms of the given reflection situation. 7) FDMA-Far means using the CSI of the farthest IRS user to optimize the phase shifts, and then each subchannel optimizes its own transmit vector in terms of the given reflection situation. 8) FDMA-EUS means exhaustive user selection to find the best user, which is selected for the reflection optimization, and then each subchannel optimizes its own transmit vector in terms of the given reflection situation. 9) FDMA-RU means randomly selecting a user for the reflection optimization, and then each subchannel optimizes its own transmit vector.  It is observed in the simulations that three iterations are sufficient for the convergence of alternating optimization.

\figurename \ref{Fig_results1} compares the cumulative distribution functions (CDFs) of the sum spectral efficiency in an IRS-aided multi-user MIMO system with $K=2$ users, consisting of a far user and a near user.  The conventional FDMA achieves the $95\%$-likely spectral efficiency (a common measure for cell-edge performance) of around \SI{4.3}{\bps\per\hertz^{}}, and $50\%$-likely or median spectral efficiency of  \SI{7.6}{\bps\per\hertz^{}}. As expected, TDMA has the exactly same performance as FDMA in the absence of IRS. NOMA is superior to FDMA/TDMA due to the superposition of more users over the same time-frequency resource, where the $95\%$-likely and $50\%$-likely rates are increased to approximately \SI{5.3}{\bps\per\hertz^{}} and \SI{9.7}{\bps\per\hertz^{}}, respectively. It is
verified that the use of IRS brings substantial performance improvement.  Even though random phase shifts are applied, the $95\%$-likely and $50\%$-likely rates of FDMA-RP grow to \SI{15}{\bps\per\hertz^{}} and \SI{18.6}{\bps\per\hertz^{}}, respectively, amounting to almost three-fold improvement compared to FDMA without the aid of IRS. Due to exhaustive best-user searching and joint reflection optimization, FDMA-EUS further boosts the performance, reaching
\SI{19}{\bps\per\hertz^{}} and \SI{22.5}{\bps\per\hertz^{}}. Interestingly, selecting either a near or far user suffers from a slight rate loss. That is to say, user scheduling does not provide a meaningful gain in the FDMA-IRS system. Compared with TDMA-IRS, achieving  $95\%$-likely spectral efficiency of \SI{22.18}{\bps\per\hertz^{}} and $50\%$-likely spectral efficiency of \SI{25.81}{\bps\per\hertz^{}}, there is a loss of approximately \SI{3}{\bps\per\hertz^{}}. That is because the time-selective reflection optimally optimizes both users in TDMA-IRS, whereas one FDMA user suffers from phase-unaligned reflection due to the lack of frequency-selective reflection.

In addition, we also illustrate the performance comparison of these schemes in the case of $K=20$ users. We observe that either random user scheduling or random phase reflection achieves the near-optimal performance, which has less than \SI{1}{\bps\per\hertz^{}} rate loss, in comparison with exhaustive user selection and optimal reflection optimization. That is because only one user gets the perfect performance with the aid of the IRS, whereas most of the users ($K-1$) suffer from un-aligned phase reflection. With the increasing number of users, the preference is random user selection and random phase shifts, which lowers the system complexity substantially but achieves the near-optimal performance.

\section{Conclusions}
This paper investigated multi-user scheduling and passive-beamforming optimization for (O)FDMA in a multi-user IRS communications system. For a comprehensive view, different user-scheduling strategies including exhaustive search, random search, farthest user, and nearest user, as well as different passive-beamforming methods, i.e., joint reflection optimization and random reflection, were presented. Theoretical analysis and numerical evaluation revealed a meaningful outcome. That is, FDMA-IRS does not need user scheduling when there are a few users. If the number of users becomes large, neither user scheduling nor reflection optimization is needed. This findings helps substantially simplifying the design of IRS-aided (O)FDMA communications systems (e.g., the estimation of cascaded channel is avoided) but achieve the near-optimal performance.





%

\bibliographystyle{IEEEtran}
\bibliography{IEEEabrv,Ref_GCWS}

\begin{thebibliography}{10}
\providecommand{\url}[1]{#1}
\csname url@samestyle\endcsname
\providecommand{\newblock}{\relax}
\providecommand{\bibinfo}[2]{#2}
\providecommand{\BIBentrySTDinterwordspacing}{\spaceskip=0pt\relax}
\providecommand{\BIBentryALTinterwordstretchfactor}{4}
\providecommand{\BIBentryALTinterwordspacing}{\spaceskip=\fontdimen2\font plus
\BIBentryALTinterwordstretchfactor\fontdimen3\font minus
  \fontdimen4\font\relax}
\providecommand{\BIBforeignlanguage}[2]{{%
\expandafter\ifx\csname l@#1\endcsname\relax
\typeout{** WARNING: IEEEtran.bst: No hyphenation pattern has been}%
\typeout{** loaded for the language `#1'. Using the pattern for}%
\typeout{** the default language instead.}%
\else
\language=\csname l@#1\endcsname
\fi
#2}}
\providecommand{\BIBdecl}{\relax}
\BIBdecl

\bibitem{Ref_liu2021reconfigurable}
Y.~Liu \emph{et~al.}, ``Reconfigurable intelligent surfaces: Principles and
  opportunities,'' \emph{IEEE Commun. Surveys Tuts.}, vol.~23, no.~3, pp. 1546
  -- 1577, Third Quarter 2021.

\bibitem{Ref_wu2019intelligent}
Q.~Wu and R.~Zhang, ``Intelligent reflecting surface enhanced wireless network
  via joint active and passive beamforming,'' \emph{{IEEE} Trans. Wireless
  Commun.}, vol.~18, no.~11, pp. 5394 -- 5409, Nov. 2019.

\bibitem{Ref_renzo2020reconfigurable}
M.~D. Renzo \emph{et~al.}, ``Reconfigurable intelligent surfaces vs. relaying:
  Differences, similarities, and performance comparison,'' \emph{IEEE Open J.
  Commun. Society}, vol.~1, pp. 798 -- 807, Jun. 2020.

\bibitem{Ref_jiang2021road}
W.~Jiang \emph{et~al.}, ``The road towards {6G}: A comprehensive survey,''
  \emph{IEEE Open J. Commun. Society}, vol.~2, pp. 334--366, Feb. 2021.

\bibitem{Ref_wang2020channel}
Z.~Wang, L.~Liu, and S.~Cui, ``Channel estimation for intelligent reflecting
  surface assisted multiuser communications: Framework, algorithms, and
  analysis,'' \emph{{IEEE} Trans. Wireless Commun.}, vol.~19, no.~10, pp. 6607
  -- 6620, Oct. 2020.

\bibitem{Ref_zhi2021uplink}
K.~Zhi \emph{et~al.}, ``Uplink achievable rate of intelligent reflecting
  surface-aided millimeter-wave communications with low-resolution {ADC} and
  phase noise,'' \emph{IEEE Wireless Commun. Lett.}, vol.~10, no.~3, pp. 654 --
  658, Mar. 2021.

\bibitem{Ref_zheng2020intelligent}
B.~Zheng and R.~Zhang, ``Intelligent reflecting surface-enhanced {OFDM}:
  Channel estimation and reflection optimization,'' \emph{IEEE Wireless Commun.
  Lett.}, vol.~9, no.~4, pp. 518 -- 522, Apr. 2020.

\bibitem{Ref_hu2018beyond}
S.~Hu, F.~Rusek, and O.~Edfors, ``Beyond massive {MIMO}: The potential of data
  transmission with large intelligent surfaces,'' \emph{{IEEE} Trans. Signal
  Process.}, vol.~66, no.~10, pp. 2746 -- 2758, May 2018.

\bibitem{Ref_di2020hybrid}
B.~Di \emph{et~al.}, ``Hybrid beamforming for reconfigurable intelligent
  surface based multi-user communications: Achievable rates with limited
  discrete phase shifts,'' \emph{{IEEE} J. Sel. Areas Commun.}, vol.~38, no.~8,
  pp. 1809 -- 1822, Aug. 2020.

\bibitem{Ref_jiang2022dualbeam}
W.~Jiang and H.~Schotten, ``Dual-beam intelligent reflecting surface for
  millimeter and {THz} communications,'' in \emph{Proc. 2022 IEEE 95th Veh.
  Techno. Conf. (VTC2022-Spring)}, Helsinki, Finland, Jun. 2022.

\bibitem{Ref_jiang2016ofdm}
W.~Jiang and T.~Kaiser, ``From {OFDM} to {FBMC}: Principles and
  {Comparisons},'' in \emph{Signal Processing for 5G: Algorithms and
  Implementations}, F.~L. Luo and C.~Zhang, Eds.\hskip 1em plus 0.5em minus
  0.4em\relax United Kindom: John Wiley\&Sons and IEEE Press, 2016, ch.~3.

\bibitem{Ref_zheng2020intelligent_COML}
B.~Zheng, Q.~Wu, and R.~Zhang, ``Intelligent reflecting surface-assisted
  multiple access with user pairing: {NOMA or OMA}?'' \emph{{IEEE} Commun.
  Lett.}, vol.~24, no.~4, pp. 753 -- 757, Apr. 2020.

\bibitem{Ref_chen2019toward}
Y.~Chen \emph{et~al.}, ``Toward the standardization of non-orthogonal multiple
  access for next generation wireless networks,'' \emph{{IEEE} Commun. Mag.},
  vol.~56, no.~3, pp. 19 -- 27, Mar. 2018.

\bibitem{Ref_chu2022resource}
Z.~Chu \emph{et~al.}, ``Resource allocation for {IRS}-assisted wireless-powered
  {FDMA IoT} networks,'' \emph{IEEE Internet of Things J.}, vol.~9, no.~11, pp.
  8774 -- 8785, Jun. 2022.

\bibitem{Ref_renzo2020smart}
M.~D. Renzo \emph{et~al.}, ``Smart radio environments empowered by
  reconfigurable intelligent surfaces: How it works, state of research, and the
  road ahead,'' \emph{{IEEE} J. Sel. Areas Commun.}, vol.~38, no.~11, pp. 2450
  -- 2525, Nov. 2020.

\bibitem{Ref_wu2020beamforming}
Q.~Wu and R.~Zhang, ``Beamforming optimization for wireless network aided by
  intelligent reflecting surface with discrete phase shifts,'' \emph{{IEEE}
  Trans. Commun.}, vol.~68, no.~3, pp. 838 -- 1851, Mar. 2020.

\bibitem{Ref_jiang2022impact}
W.~Jiang and H.~D. Schotten, ``Impact of channel aging and phase noise on
  intelligent reflecting surface,'' \emph{{IEEE} Commun. Lett.}, 2022,
  submitted.

\bibitem{Ref_jiang2021cellfree}
------, ``Cell-free massive {MIMO-OFDM} transmission over frequency-selective
  fading channels,'' \emph{{IEEE} Commun. Lett.}, vol.~25, no.~8, pp. 2718 --
  2722, Aug. 2021.

\end{thebibliography}

\end{document}